\documentclass[aps,preprint,a4paper,showpacs,showkeys,onecolumn,12pt]{revtex4}%
\usepackage{amsfonts}
\usepackage{amsmath}
\usepackage{amssymb}
\usepackage{graphicx}%
\begin{document}
\title{From Darwin to Sommerfeld: Genetic algorithms and the electron gas}
\author{C\'{e}sar O. Stoico}
\affiliation{Area F\'{\i}sica, Facultad de Ciencias Bioqu\'{\i}micas y Farmac\'{e}uticas,
Universidad Nacional de Rosario, Argentina.}
\author{Danilo G. Renzi}
\affiliation{Facultad de Ciencias Veterinarias, Universidad Nacional de Rosario, Casilda, Argentina}
\author{Fernando Vericat}
\affiliation{Instituto de F\'{\i}sica de L\'{\i}quidos y Sistemas Biol\'{o}gicos
(IFLYSIB)-CONICET-UNLP, 59 789 c.c. 565 (1900) , La Plata, Argentina.}
\altaffiliation{Also at Grupo de Aplicaciones Matem\'{a}ticas y Estad\'{\i}sticas de la
Facultad de Ingenier\'{\i}a (GAMEFI), UNLP, La Plata, Argentina}

\altaffiliation{E-mail: vericat@iflysib.unlp.edu.ar}

\begin{abstract}
In return for the long-standing contributions of Physics to Biology, now the
inverse way is frequently traveled through in order to think about many
physics phenomena. In this vein, evolutionary algorithms, particularly genetic
algorithms, are being more and more used as a tool to deal with several
Physics problems. Here, we show how to apply a genetic algorithm to describe
the homogeneous electron gas.

\end{abstract}
\maketitle

Several methods, techniques and ideas taken from Physics have been of primary
importance for the notable development of Biology in the last sixty years. To
agree with this statement is it enough to think about the application of X Ray
crystallography to the resolution of protein tertiary and quaternary
structures, or the fundamental role that Quantum Mechanics in general and
Molecular Physics in particular have played in the establishment of modern
(Molecular) Biology, as predicted by Schr\"{o}dinger\cite{Schroedinger1}.
However, since some years ago, the inverse way is also frequently traveled by
physicists. Biological methods and concepts, as well as its terminology
itself, are more and more used in Physics.

In this context, in the last decades, there has been an increasing interest in
algorithms based on the Darwinian evolution principle\cite{Holland1}. Genetic
algorithms\cite{Goldberg1}-\cite{Mitchell1}, evolutionary
programming\cite{Fogel1},\cite{Back1}, game-playing strategies\cite{Davis1}
and genetic programming\cite{Koza1} have found a wide field of applications,
not just in Physics but also in other areas where optimization plays an
important role, such as financial markets, artificial intelligence, etc. In
particular, genetic algorithms tackle even complex problems with surprising
efficiency and robustness. In Physics they have been used in calculations thet
involve from single Schr\"{o}dinger particles in diverse potentials to
astrophysical systems, running through lattice systems, spin glass models,
molecules and clusters. The differences among the various evolutionary
algorithms can be found not in the basic principles but in the details of the
selection, reproduction and mutation procedures.

In general, an evolutionary algorithm is based on three main statements:

a) It is a process that works at the chromosomic level. Each individual is
codified as a set of chromosomes.

b) The process follows the Darwinian theory of evolution, say, the survival
and reproduction of the fittest in a changing environment.

c) The evolutionary process takes place at the reproduction stage. It is in
this stage when mutation and crossover occurs. As a result, the progeny
chromosomes can differ from their parents ones.

Starting from a guess initial population, an evolutionary algorithm basically
generates consecutive generations (offprints). These are formed by a set of
chromosomes, or character (genes) chains, which represent possible solutions
to the problem under consideration. At each algorithm step, a fitness function
is applied to the whole set of chromosomes of the corresponding generation in
order to check the goodness of the codified solution. Then, according to their
fitting capacity, couples of chromosomes, to which the crossover operator will
be applied, are chosen. Also, at each step, a mutation operator is applied to
a number of randomly chosen chromosomes.

The two most commonly used methods to randomly select the chromosomes are:

i) \textit{The roulette wheel algorithm}. It consists in building a roulette,
so that to each chromosome corresponds a circular sector proportional to its fitness.

ii) \textit{The tournament method}. After shuffling the population, their
chromosomes are made to compete among them in groups of a given size
(generally in pairs). The winners will be those chromosomes with highest
fitness. If we consider a binary tournament, say the competition is between
pairs, the population must be shuffled twice. This technique guarantees copies
of the best individual among the parents of the next generation.

After this selection, we proceed with the sexual reproduction or crossing of
the chosen individuals. In this stage, the survivors exchange chromosomic
material and the resulting chromosomes will codify the individuals of the next
generation. The forms of sexual reproduction most commonly used are:

i) With one crossing point. This point is randomly chosen on the chain length,
and all the chain portion between the crossing point and the chain end is exchanged.

ii)\ With two crossing points. The portion to be exchanged is in between two
randomly chosen points.

For the algorithm implementation, the crossover normally has an assigned
percentage that determines the frequency of its occurrence. This means that
not all of the chromosomes will exchange material but some of them will pass
intact to the next generation. As a matter of fact, there is a technique,
named elitism, in which the fittest individual along several generations does
not cross with any of the other ones and keeps intact until an individual
fitter than itself appears.

Besides the selection and crossover, there is another operation, mutation,
that produces a change in one of the\ characters or genes of a randomly chosen
chromosome. This operation allows to introduce new chromosomic material into
the population. As for the crossover, the mutation is handled as a percentage
that determines its occurrence frequency. This percentage is, generally, not
greater than 5\%, quite below the crossover percentage.

Once the selected chromosomes have been crossed and muted, we need some
substitution method. Namely,\ \ we must choose, among those individuals, which
ones will be substituted for the new progeny. Two main substitution ways are
usually considered. In one of them, all modified parents are substituted for
the generated new individuals. In this way an individual does never coexist
with its parents. In the other one, only the worse fitted individuals of the
whole population are substituted, thus allowing the coexistence among parents
and progeny.

Since the answer to our problem is almost always unknown, we must establish
some criterion to stop the algorithm. We can mention two such criteria: i) the
algorithm is run along a maximum number of generations; ii) the algorithm is
ended when the population stabilization has been reached, i.e. when all, or
most of, the individuals have the same fitness.

In this letter we apply some of the previous ideas in order to calculate the
pair distribution function (PDF) of an homogeneous electron gas. For clarity
we consider the one-dimensional version of the electron gas, but the method is
directly extended to higher dimensions\cite{Stoico1}.

Let us consider the Sommerfeld-Pauli model of metallic solids\cite{Hoddeson1}
in 1D, namely $N$ electrons of mass $m$ moving along a segment of the axis $x$
of length $L$ . The position of the $i$th electron is denoted with $x_{i}$
($i=1,2,...,N$). Electrons $i$ and $j$ interact one with the other through a
pair potential $v\left(  x_{ij}\right)  =\left(  x_{ij}^{2}+\delta^{2}\right)
^{-1/2}$ . This pair potential is often used to model a quantum wire of width
$\delta$. The system Hamiltonian then reads:%

\begin{equation}
H=%
%TCIMACRO{\dsum \limits_{i}^{N}}%
%BeginExpansion
{\displaystyle\sum\limits_{i}^{N}}
%EndExpansion
-\frac{\hbar^{2}}{2m}\nabla_{i}^{2}+%
%TCIMACRO{\dsum \limits_{i<j}^{N}}%
%BeginExpansion
{\displaystyle\sum\limits_{i<j}^{N}}
%EndExpansion
v\left(  x_{ij}\right)  . \tag{1}\label{1}%
\end{equation}

For the $N$-body wave function we use the trial form\cite{Lado1}%

\begin{equation}
\Psi\left(  \overrightarrow{x}\right)  =\exp\left[  -%
%TCIMACRO{\dsum \limits_{i<j}^{N}}%
%BeginExpansion
{\displaystyle\sum\limits_{i<j}^{N}}
%EndExpansion
u\left(  x_{ij}\right)  +%
%TCIMACRO{\dsum \limits_{i<j}^{N}}%
%BeginExpansion
{\displaystyle\sum\limits_{i<j}^{N}}
%EndExpansion
w\left(  x_{ij}\right)  \right]  , \tag{2}\label{2}%
\end{equation}
where $\overrightarrow{x}\equiv\left(  x_{1},x_{2},...,x_{N}\right)  $ denotes
the system configuration and%

\begin{equation}
w\left(  x\right)  =\ln g_{0}\left(  x\right)  -\frac{1}{2\pi\rho}%
%TCIMACRO{\dint }%
%BeginExpansion
{\displaystyle\int}
%EndExpansion
dke^{-ikx}\frac{\left[  \tilde{S}_{0}\left(  k\right)  -1\right]  ^{2}}%
{\tilde{S}_{0}\left(  k\right)  }. \tag{3}\label{3}%
\end{equation}
Here $\rho=N/L$ is the system number density and\ $g_{0}\left(  x\right)  $ is
the ideal ($v\left(  x\right)  \equiv0$) electron gas pair distribution
function with its Fourier transform $\tilde{S}_{0}\left(  k\right)  $, the
structure factor.

To apply the algorithm, we take $M$ distances $y_{1}=0,$ $y_{2}=d,...,y_{M}%
=(M-1)d$ so that, for any $i$,$j$, the function $u\left(  x_{ij}\right)  $ is
represented by the random string $u\left(  x_{ij}=y_{1}\right)  $, $u\left(
x_{ij}=y_{2}\right)  $,...,$u\left(  x_{ij}=y_{M}\right)  $, where each
$u\left(  x_{ij}=y_{k}\right)  $ is a random real number $\gamma_{ij}^{k}%
\in\left[  0,1\right]  $ (rounded to an established number $n$ of decimals):
$u\left(  x_{ij}=y_{k}\right)  =\gamma_{ij}^{k}$. Our initial population is
then formed by $N_{p}$ random replicas of each of the $N\left(  N-1\right)
/2$ strings. In this work we have taken $N=100$, $M=99$, $n=3$ and $N_{p}=100$.

Given a string $\gamma_{ij}^{1},\gamma_{ij}^{2},...,\gamma_{ij}^{M}$, the
\textit{encoding }consists\textit{ }in replacing the sequence of \ real
numbers $\gamma_{ij}^{k}\in\left[  0,1\right]  $ by a single natural number
obtained by putting their decimal parts one next to the other. For example,
taking $n=3$: the string $0.137,0.935,...,0.466$ gives $137935...466$. In
genetic terms, the encoding produces the chromosomic structures of the parents
(strings). The inverse process is called \textit{decoding}.

For the parents population, the energy of the $\alpha$th replica
($\alpha=1,2,...N_{p}$) is given by%

\begin{equation}
E_{\alpha}=\frac{\left\langle \Psi_{\alpha}\right\vert H\left\vert
\Psi_{\alpha}\right\rangle }{\left\langle \Psi_{\alpha}\right\vert \left.
\Psi_{\alpha}\right\rangle }, \tag{4}\label{4}%
\end{equation}
where for the mean kinetic energy we use the Jackson-Feenberg
formulae\cite{Feenberg1}.

In order to calculate the fitness of the replica $\alpha$, we previously need
to estimate the energy error: $\epsilon_{\alpha}=\left\vert \left(
H-E_{\alpha}\right)  \Psi_{\alpha}\right\vert $ and define the fitness
function as $f_{\alpha}=e^{-\epsilon_{\alpha}}$. A solution is reached when
$f_{\alpha}\approx1$.

The calculation proceeds by dividing the population of $N_{p}$ replicas into
$N_{p}/2$ couples. The couples are randomly chosen by using the roulette wheel
algorithm\cite{Golberg1}. This is done by defining the sums $F=%
%TCIMACRO{\tsum \nolimits_{\alpha=1}^{N_{p}}}%
%BeginExpansion
{\textstyle\sum\nolimits_{\alpha=1}^{N_{p}}}
%EndExpansion
f_{\alpha}$ and $S_{\beta}=%
%TCIMACRO{\tsum \nolimits_{\alpha=1}^{\beta}}%
%BeginExpansion
{\textstyle\sum\nolimits_{\alpha=1}^{\beta}}
%EndExpansion
$ $f_{\alpha}$ \ ($\beta=1,2,...,N_{p}$). Then, a random number $r\in\left[
0,F\right]  $ is generated and the unique index $\delta$ such that
$S_{\delta-1}\leq r\leq S_{\delta\text{ }}$is picked up.

Once the first generation of replicas (parents) has been generated and divided
into couples, the second generation (offspring) can be generated by applying
the crossover operator between the members of each one. Sometimes, some of the
members of the new replicas generation can be changed by applying the mutation operator.

Given a couple of replicas, the crossover operator is defined by generating a
new random number $c\in\left[  0,1\right]  $ which is compared with a
pre-established crossover probability $p\in\left[  0,1\right]  $. If $c\leq
p$, the crossover operator acts by interchanging all the digits from the $s$th
position to the end of the replica between the members of the couple.\ Here
$s$ is a random integer such that $1\leq s\leq nM$. for example if the couple is

$153280472...337$

$768325399...069$

and $s=4,$ the new offspring couple will be

$153225399...069$

$768380472...337.$

To apply the mutation operator we first randomly select those offsprings that
will mutate. Then, for each of these offsprings, a gene (a digit) randomly
chosen is changed by a random integer number $\ell\in\left[  0,9\right]  .$

The algorithm was stopped after $2\times10^{4}$ generations. Then, the PDF at
the distances $y_{1},y_{2},...,y_{M}$ is calculated%

\begin{equation}
g\left(  x\right)  =\frac{N(N-1)}{\rho^{2}}\frac{\int dx_{3}dx_{4}\ldots
dx_{N}\left\vert \Psi(x_{1},x_{2},\ldots x_{N})\right\vert ^{2}}{\int
dx_{1}dx_{2}\ldots dx_{N}\left\vert \Psi(x_{1},x_{2},\ldots x_{N})\right\vert
^{2}}. \tag{5}\label{5}%
\end{equation}

The so-calculated PDF$^{\prime}$s were compared with those
obtained\cite{Comment1} from Fermi hypernetted chain (FHNC) and Monte Carlo
(MC) variational approaches with good agreement for different values of the
relevant parameters (Figs. 1-3). FHNC and MC probably are the two most
reliable methods available to calculate the electron gas correlations.

Taking into account the good agreement among the curves, and the above
mentioned efficiency and robustness of the genetic algorithm, we conclude that
the application of this kind of algorithms to many-body problems deserves some
attention from physicists in order to achieve a better understanding of the
involved physical subtleties\cite{Stoico1}.

\begin{acknowledgments}
Support of this work by Universidad Nacional de Rosario (UNR), Universidad
Nacional de La Plata (UNLP), Consejo Nacional de Investigaciones
Cient\'{\i}ficas y T\'{e}cnicas (CONICET) and Agencia Nacional de
Promoci\'{o}n Cient\'{\i}fica y Tecncnol\'{o}gica\ (ANPCyT) of Argentina is
greatly appreciated. F.V. is a member of CONICET.
\end{acknowledgments}

\begin{center}
\textbf{Figure Captions}
\end{center}

\textbf{Figure 1. }The\textbf{ }pair distribution function for the 1D
homogenous electron gas with Wigner-Seitz radius \textit{r}$_{s}$=1 and
screening parameter $\delta$=1.

\textbf{Figure 2. }Same as Fig. 1 for \textit{r}$_{s}$=1 and $\delta$=2.

\textbf{Figure 3. }Same as Fig. 1 for \textit{r}$_{s}$=4.5 and $\delta$=1.


\begin{thebibliography}{99}                                                                                               %


\bibitem {Schroedinger1}E. Schr\"{o}dinger, \textit{What is life?} \textit{The
Physical Aspect of Living Cell }(Cambridge University Press, Cambridge, 1944).

\bibitem {Holland1}J.H. Holland, \textit{Adaptation in Natural and Artificial
Systems} (University of Michigan, Ann Arbor, MI, 1975).

\bibitem {Goldberg1}D.E. Goldberg, \textit{Genetic Algorithm in Search,
Optimization and Machine Learning} (Addison-Wesley, Reading, MA, 1989).

\bibitem {Mitchell1}M. Mitchell, \textit{An Introduction to Genetic Algorithm}
(Prentice Hall, 1998).

\bibitem {Fogel1}D. Fogel, \textit{Evolutionary Computation} (IEEE Press, 1996).

\bibitem {Back1}T. B\"{a}ck, \textit{Evolutionary Algorithm in Theory and
Practice} (Oxford Press, 1996).

\bibitem {Davis1}\textit{Genetic Algorithms and Simulated Annealing}, edited
by L. Davis (Pitman. London. 1987)-

\bibitem {Koza1}J.R. Koza, \textit{Genetic Programming: On the Programming of
Computers by Means of Natural Selection }(MIT, Cambridge, MA,1992).

\bibitem {Stoico1}C.O. Stoico and F. Vericat, \textit{Genetic algorithm and
the electron gas in 1, 2 and 3D }(in preparation).

\bibitem {Hoddeson1}L. Hoddeson, G. Baym and M. Eckert, Rev. Mod. Phys.
\textbf{59}, 287 (1987).

\bibitem {Lado1}F. Lado, J. Chem. Phys. \textbf{47}, 5369 (1967).

\bibitem {Feenberg1}Feenberg, \textit{Theory of Quantum Fluids }(Academic
Press, New York, 1969).

\bibitem {Comment1}See reference \cite{Stoico1} for details.\newpage
\end{thebibliography}
\end{document}